# How Spontaneous Electrowetting and Surface Charge affect Drop Motion

Chirag Hinduja[1*], Benjamin Leibauer[1], Rishi Chaurasia[1,3], Nikolaus Knorr[1], Aaron D. Ratschow[1,2], Shalini Singh[1,4], Hans-Jürgen Butt[1*], Rüdiger Berger[1*]

[1] Max Planck Institute for Polymer Research, 55128 Mainz, Germany

[2] Institute for Nano- and Microfluidics, TU Darmstadt, 64287 Darmstadt, Germany

[3] Indian Institute of Science Education and Research, Tirupati-517619, India

[4] Indian Institute of Technology Delhi, Hauz Khas, New Delhi-110016, India

[*]Corresponding Authors; E-mail: butt@mpip-mainz.mpg.de; berger@mpip-mainz.mpg.de; hindujachirag604@gmail.com

## Abstract

Water drops sliding on hydrophobic surfaces spontaneously separate charges at their rear. It is unclear how this charge separation affects the contact angles of a sliding drop. We slide grounded and insulated drops on hydrophobic surfaces at low capillary numbers ($\leq 10^{-4}$). We find that drop charge leads to spontaneous electrowetting which decreases the contact angles. Additionally, the deposited charges lead to surface charge effect and decrease the contact angle. Both phenomena compensate each other at the receding contact line, resulting in an insignificant change in receding contact angle of a sliding drop.

# Introduction

Drop sliding is a universal phenomenon and spans multiple disciplines. For instance, the back surface of a desert beetle inspires research into the development of water harvesting technologies[1-3], drop-wise condensation for enhanced heat transfer[4], retention of agricultural sprays on plants leaves[5], and drug production using droplet microfluidics[6]. To achieve the desired wetting properties, the surface energy is lowered by employing a water repellent coating. When a water drop slides on a hydrophobized surface, charges separate at the receding contact line and the drop acquires a net charge[7-14]. This phenomenon is called *slide electrification* or *solid-liquid contact electrification*[9,10,13,15]. Slide electrification alters the governing forces on the drop and influences drop motion[16]. By sliding grounded and insulated drops at low capillary numbers, we investigate the influence of slide electrification on the sliding contact angles.

Slide electrification of drops can lead to two phenomena, which decrease the contact angles of a sliding drop. The first is electrowetting. The drop charge causes an electric potential $U$ with respect to the ground electrode beneath the substrate[17,18]. Integrating the energy of the electric field across the thickness ($d$) of the dielectric material and assuming that the drop size is much large than $d$ leads to an energy per unit area of $\varepsilon_o \varepsilon_s U^2/2d$.[19-22] Formally, this can be viewed as a reduction in the solid-liquid interfacial energy ($\Delta \gamma_{SL}$). A decrease in solid-liquid interface energy in turn decreases the advancing contact angle ($\theta_a$) at the front and receding contact angle ($\theta_r$) at the rear of the sliding drop (Fig 1a). The decrease in $\theta_a$ and $\theta_r$ is described by.

$$\cos\theta_{a/r}(U) - \cos\theta_{a/r}(U=0) = -\frac{\Delta\gamma_{SL}}{\gamma} = \frac{\varepsilon_o \varepsilon_s}{2d} \cdot \frac{U^2}{\gamma} \quad (1)$$

Here, $\theta_{a/r}(U=0)$ and $\theta_{a/r}(U)$ are the advancing and receding contact angles (CAs) in the absence and presence of a drop potential $U$, respectively, $\gamma$ is surface tension of the liquid, $\varepsilon_s$ and $\varepsilon_o$ are the dielectric permittivities of the substrate and vacuum, respectively. In typical electrowetting experiments, $U$ is applied to the drop by an external electrode. However, in slide electrification, the charges separate spontaneously during the drop motion. Hereafter, we term this phenomenon *spontaneous electrowetting*.

The second possible phenomenon in slide electrification is the change in wettability of the surface due to the deposition of charges onto the surface[23,24]. Depositing charges onto the surface requires electrostatic work to be carried out which effectively increase the solid-air interfacial energy ($\Delta\gamma_S$) which scales quadratically with the surface charge density[20] (Fig 1b). This increase in surface energy decreases the CA of the drop. Using Young's equation locally at the rear contact line yields (eq. 2).

$$\cos\theta_r(\sigma) - \cos\theta_r(\sigma=0) = \frac{\Delta\gamma_S}{\gamma} \propto \frac{\sigma^2}{\gamma} \quad (2)$$

Here, $\theta_r(\sigma)$ is the receding CA with charge separation, $\theta_r(\sigma=0)$ is the CA without charge separation, $\sigma$ is the surface charge density of the solid-air interface. The precise value of $\Delta\gamma_S$ depends on the shape and size of the charged region. We term this phenomenon the *surface charge* effect.

Both effects, *spontaneous electrowetting* and *surface charge*, depend on the charge separation. Charge separation depends on the drop velocity $v$ [25,26] and surface wetting properties[27,28]. Furthermore, they coexist with hydrodynamic viscous dissipation[29,30]. Viscous dissipation inside the drop increases the advancing CA and decreases the receding CA of a sliding drop at high capillary numbers[31-33], $Ca = \frac{\mu v}{\gamma} > 10^{-3}$ (Fig 1a,b). Here, µ is the dynamic viscosity of the liquid.

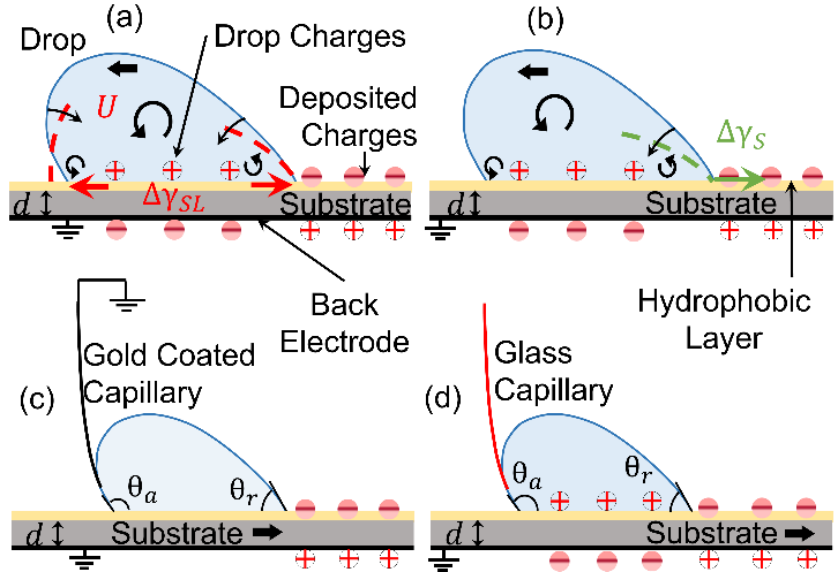

**Figure 1.** Schematic changes in CAs of a sliding drop due to (a) spontaneous electrowetting and (b) surface charge effect. Drop charge results in spontaneous electrowetting ($\Delta\gamma_{SL}$) and deposited charges increase solid surface energy ($\Delta\gamma_S$). Both effects reduce the CAs of a sliding drop (red and green dashed lines in (a) and (b), respectively) and coexist with viscous dissipation (circular arrows). The black arrow in (a, b) indicates the direction of drop motion. The thickness of hydrophobic layer is ~ nm (yellow) and substrate is ~ mm (gray). (c) Drop sliding using a gold-coated electrically conductive capillary. It constantly discharges the drop and maintains it at zero electric potential. (d) Drop sliding using a glass capillary. It results in charges staying in the drop. The black arrow indicates the direction of stage motion.

All three phenomena, spontaneous electrowetting, surface charge, and viscous dissipation affect the contact angles and determine the forces acting on the sliding drop. However, it is unclear to what extent slide electrification contributes to the contact angles of a sliding drop. In this article, we separate and quantify spontaneous electrowetting and surface charge effects in the absence of viscous dissipation. Both effects need to be considered for a complete understanding of the physics of drop motion.

In contrast to tilted plate experiments[9,28,34,35], we mitigate viscous dissipation by sliding a 5 µL milli-Q water drop at constant velocities in the range 0.1—10 mm/s ($Ca \sim 10^{-6}$—$10^{-4}$)[36-39]. Sliding a drop using an electrically conductive capillary results in a grounded drop, i.e., the drop slides at zero electric potential (Fig 1c). This allows us to omit spontaneous electrowetting[40]. As a result, only the deposited charges behind the drop affect drop sliding motion and allow us to investigate the surface charge effect. In contrast, by using a glass capillary for a sliding drop, we investigate the combined effects of *spontaneous electrowetting* and *surface charge* (Fig 1d).

## Methods

We coat Trichloro(1H,1H,2H,2H-perfluorooctyl)silane (PFOTS, 97% Alfa Aesar) and Trichloro(octyl)silane (OTS, Sigma Aldrich 97%) on 76.2 × 25.4 × 1 (or 0.25, 0.5, 2, 5) mm³ fused quartz (Thermo Fisher Scientific, resistivity ≈ $10^{15}$ Ω cm, $\varepsilon_s$ ≈ 3.5) using chemical vapor deposition process (S1)[41]. For simplicity, fused quartz substrate will be termed as *quartz* hereafter. We characterize the surfaces using surface energy measurements (S2), scanning force microscopy (S3), and the sessile drop method (S4). We measure sliding CAs for 5 µL milli-Q water drops (Sartorius, resistivity ≈ 18.2 MΩ cm, $\gamma$ = 72 mN/m). In all measurements, we slide drops for 5 cm. The substrates are placed on a metal plate, which is connected to ground. The CAs are recorded by a side-view camera[36-40,42] (Fig 1c, d). To determine advancing (leading) and receding (trailing) CAs of the drops, a tangent is fit to 10 pixels (image resolution = 6 µm/pixel) above the baseline[43,44]. To avail conductive capillary, the glass capillary is sputter-coated with 5 nm thick chromium followed by 30 nm thick gold layer (S5).

## Results

Sliding a grounded drop deposits charges at zero potential, in parallel counter charges in the grounded drop are continuously neutralized. The deposited charges at the surface increase the solid surface energy ($\Delta\gamma_S$). For a water drop sliding at 2 mm/s on a neutral (or unwetted) PFOTS-quartz, we measure advancing and receding CAs of 120° ± 2° and 60° ± 3°, respectively. Both CAs stay constant after a sliding length ≈ 4 mm and do not change in the kinetic regime of drop sliding (Fig 2a, black curve). For an electrically insulated 5 µL water drop sliding on the same neutral PFOTS-quartz, we observe that the advancing CA decreases to 113° after 15 mm stage motion. Thereafter it remains constant at 113° ± 2° along the drop sliding path (Fig 2a, red curve). We attribute the decrease in advancing CA to charging of the sliding drop. As the drop starts sliding, it deposits charges leading to a surface charge density $\sigma(x)$ at the dewetted surface. Here, $x$ is a coordinate along the drop path. The charge in the drop is equal in magnitude but opposite in polarity to the deposited charges. The accumulated charges in the drop develop an electric potential $U$, which increases according to

$$U = -\frac{w}{C}\int_0^L \sigma \, dx \qquad (3)$$

Here, $w$ is the width of the drop, $C$ is its capacitance, and $L$ is the slide length. As $U$ increases, less and less charges are deposited at the surface. After a slide length of ≈ 15 mm, the drop saturates with charges and no counter charges are deposited at the dewetted surface. This length is a direct measurement of the saturation length ($\lambda$) which is not directly measurable on tilted plate experiments[9,13]. This value which agrees with estimated saturation lengths in literature[18].

Once the drop saturates with charges, we measure a difference of 7° between the advancing CA of the grounded and the charge-saturated drop (Fig 2a). Since there are no charges on PFOTS-quartz in front of the drop, the decrease in advancing CA of an insulated drop is due to spontaneous electrowetting (left side, Fig 2b). Using eq. (1) we calculate a drop potential of ≈ 700 V. It is of the same order as reported for drops sliding down a tilted plane[8,17,18]. This spontaneous drop potential corresponds to a decrease of 8 ± 3 mN/m in solid-liquid interfacial energy ($\Delta\gamma_{SL}$). The error represents variation along the drop path.

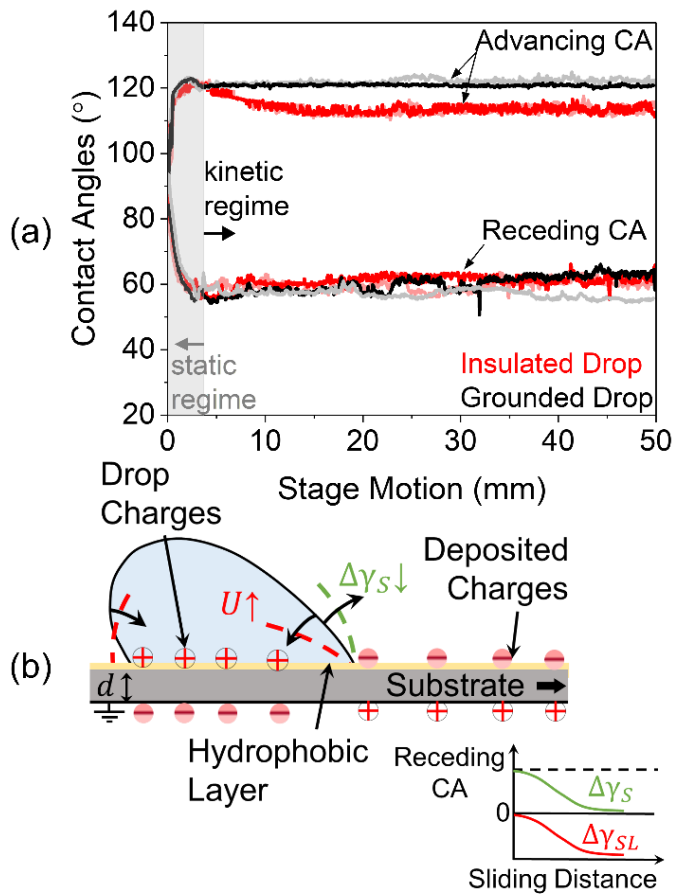

**Figure 2.** (a) CA for grounded and insulated drops that slide on a PFOTS-quartz surface. The grounded drop is sliding at zero potential, while the insulated drop accumulates charges. The black/gray data correspond to two independent measurements for grounded drops and red/pink corresponds to two independent measurements for insulated drops of 5 μL milli-Q water drops sliding at 2 mm/s. (b) The advancing CA of an insulated drop decreases due to spontaneous electrowetting, sketched as the change in black drop profile to the red dashed line. The receding CA of an insulated drop is subjected to two coupled phenomena: spontaneous electrowetting and surface charge effect. Spontaneous electrowetting decreases the receding CA, sketched as the change in black drop profile to red dashed line. Simultaneously, decrease in surface charge effect increases the receding CA, sketched as the change in black line to green dashed line. The receding CA of a grounded drop is only subjected to the surface charge effect. The black arrow indicates the direction of stage motion. Inset schematic shows a qualitative change in solid-air interfacial energy ($\Delta\gamma_S$) which is compensated by the change in solid-liquid interfacial energy ($\Delta\gamma_{SL}$). This compensation results in insignificant change in receding CA (black dashed line).

For the same insulated drop, we determine that the receding CA decreases to 60° at the onset of sliding and remains at 60° in the kinetic regime of drop sliding (Fig 2a, red curve). In particular, the receding CA for both grounded and insulated drops are similar. Due to the increasing drop potential and corresponding electrowetting effect, one would expect lower receding CA for an insulated drop compared to a grounded drop. But why is the receding CA of both grounded and insulated drop the same? Unlike at the advancing contact line, where only spontaneous electrowetting occurs, the receding contact line is subject to two simultaneous phenomena: spontaneous electrowetting and the surface charge effect. At the onset of drop motion, the drop deposits charges at zero potential and increases the solid surface energy ($\Delta\gamma_S$). As soon as the drop begins sliding, it accumulates the opposite charges. These charges in the drop develop a potential $U$. This drop potential decreases the charge separation at the receding contact line. With further contact line motion, the deposited surface charge density decreases to $\sigma \cong 0$ and the drop charge saturates. According to eq. 2 ($\Delta\gamma_S \sim \sigma^2$), a decrease in $\sigma$ leads to a decrease in $\Delta\gamma_S$, which increases the receding CA (green dashed line at receding CA, Fig 2b). At the same time, the drop potential decreases the solid-liquid interface energy according to eq. 1 ($-\Delta\gamma_{SL} \sim U^2$), which leads to a decrease in the receding CA (red dashed line at receding CA, Fig 2b). This decrease in potential scales with the surface charge density $\sigma$ (eq. 3). Therefore, according to eq. (1), $-\Delta\gamma_{SL} \sim \sigma^2$. We find that $\Delta\gamma_{SL}$ for the ungrounded drop is equal to the $\Delta\gamma_S$ for the grounded drop: $\Delta\gamma_{SL} = \Delta\gamma_S \propto \sigma^2$. Both effects, spontaneous electrowetting and surface charge compensate each other at the receding contact line (inset Fig 2b). As a result, the receding CA of 60° ± 3° for an insulated drop resembles that of a grounded drop (Fig 2a).

In typical electrowetting experiments on dielectrics, Nelson et al.[45] found that external potential leads to a decrease in the advancing and receding CAs of a sliding drop at capillary numbers $10^{-5} \leq Ca \leq 10^{-3}$. It does not alter CA hysteresis (CAH = $\theta_a - \theta_r$)[45]. However, in the kinetic regime of drop sliding at a $Ca \sim 10^{-5}$, we find that the CAH ($\theta_a - \theta_r$) reduces by 7° for an insulated drop (Fig 2). The reason for the decrease in CAH is due to a decrease in the advancing CA by spontaneous electrowetting and an insignificant change in the receding CA (Fig. 2b).

We develop a mathematical model to estimate the surface charge effect ($\Delta\gamma_S$) on the dewetted surface (S9). We assume a one-dimensional insulating layer of thickness $d$ (which extends infinitely in the other two directions) and dielectric permittivity $\varepsilon_S$ which is placed on a grounded plate. For such a system, the change in solid surface energy

$$\Delta\gamma_S = \frac{\sigma^2 d}{2\varepsilon_o \varepsilon_S} \qquad (4)$$

The insulator-air interface is charged with a uniform surface charge density $\sigma$. An additional grounded plate is placed in air at a large distance $L \gg d$. Taking quartz thickness of $d = 0.25$ mm, a relative permittivity $\varepsilon_S = 3.5$ and a surface charge density of $\sigma \approx 60$ μC/m² which is measured by Kelvin probe (Fig S5) in Eq. (4) provides an increase in surface energy ($\Delta\gamma_S$) of 15 mN/m. Therefore, our model (eq. 4) provides a good estimation for $\Delta\gamma_S$ for thin insulating substrates with a thickness smaller than the drop size.

We systematically vary the thickness of the PFOTS-quartz substrate as 0.25, 0.5, 1, 2, and 5 mm. For 5 µL drops sliding at 2 mm/s speed, we measure identical receding CA for both grounded and insulated drops on respective quartz thickness (Fig 3a). In particular, we measure identical receding CA for grounded and insulated drops on substrate thicknesses of 2 and 5 mm. For these two thicknesses, we anticipate that the parallel plate capacitor models employed in eq. 1 and 4 are not applicable. Our results imply that spontaneous electrowetting and surface charge effect compensate each other at the receding contact line for the quartz thicknesses in the range 0.25 – 5 mm.

The difference between the advancing CA of grounded and insulated drop decreases with increasing thickness of quartz (Fig 3a). Using Eq. 1, we calculate the maximum decrease in $\Delta\gamma_{SL} \approx 17$ mN/m for 0.25 and 0.5 mm thicknesses.

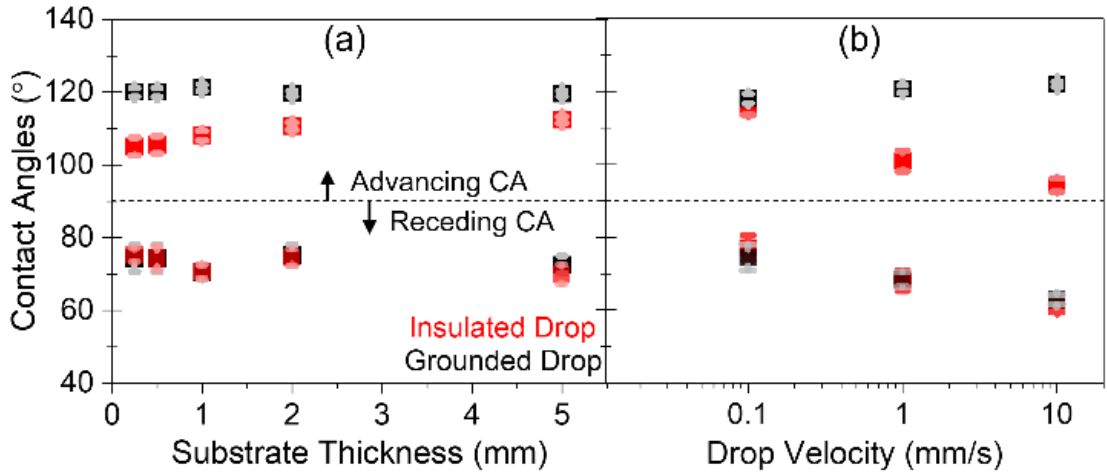

**Figure 3.** (a) Average CAs for grounded and insulated 5 µL milli-Q water drops sliding at 2 mm/s on PFOTS-quartz with thickness 0.25, 0.5, 1, 2 & 5 mm. (b) Average CAs for 5 µL water drops sliding at drop velocities of 0.1, 1, and 10 mm/s on 1 mm thick PFOTS-quartz. Stage motion of 3-5cm is considered for determining the average. Error bars correspond to variation in CAs for 3 independent drop sliding measurements.

*Influence of drop velocity.* We slide 5 µL water drops at 0.1, 1, 10 mm/s on 1 mm thick PFOTS-quartz. This velocity range corresponds to capillary numbers $10^{-6} \leq Ca \leq 10^{-4}$ for a water drop. We measure identical receding CA for both grounded and insulated drops for respective drop velocity (Fig 3b). Thus, spontaneous electrowetting and surface charge compensate at the receding contact line irrespective of the drop velocity.

For grounded drops, we observe a slight increase in advancing CA and a decrease in receding CA with increasing drop velocity (black Fig 3b). Since the charge separation for a grounded drop does not depend on drop velocity in this range[40], we attribute this observation to dissipation of the moving contact line described by the Molecular Kinetic Theory (MKT)[46, 47]. Contrary to grounded drops, the advancing CA of insulated drops decreases with increasing drop velocity (Fig 3b). We attribute this decrease in advancing CA to an increasing electric potential in the drop. With an increase in potential, the strength of spontaneous electrowetting increases (eq. 1). At lower velocity, drops may dissipate accumulated charges through the glass capillary or ambient reducing the electric potential of the drop. Furthermore, we investigate the influence of salt by varying NaCl concentration in water drops as 0.1, 1, 10, and 100 mM. For 5 µL drops sliding at 2 mm/s, no significant difference is detected in the receding CA = 89° ± 5° for grounded and insulated drops when NaCl concentration is varied from 0.1mM—100mM (Fig S6).

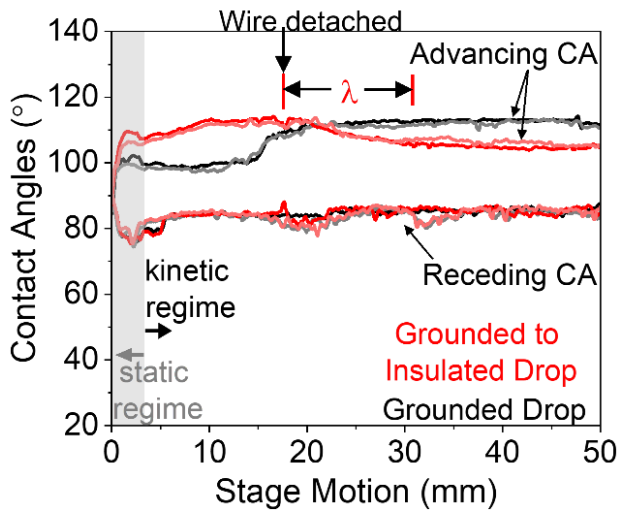

**Figure 4.** CAs of a drop when it is switched from grounded to insulated state (red). Drop grounding is achieved by attaching a conductive wire of 0.2 mm diameter to the drop. Wire detachment from the drop results in an insulated drop, indicated by a vertical arrow. CAs of grounded drops sliding along the same path as an insulated drop (black). The images correspond to a 5 µL water drop and CAs correspond to two independent 5 µL water drops sliding at 2 mm/s on a PFOTS-quartz. The CA data are smoothened by a 5 point moving average.

We extend our experiments with a glass capillary having a flexible grounded wire. Sliding a drop with a wire attached to it results in drop sliding at zero potential. The detachment of the wire from the drop switches a grounded drop to an insulated drop while the drop is in motion. A grounded 5 µL drop on a neutral PFOTS-quartz results in advancing and receding CAs of 113° and 83°, respectively (Fig 4). At about the 18 mm stage motion, we detach the wire from the drop and the drop accumulates charges. As a result, the advancing CA gradually decreases and stagnates at 105° from 32 mm of the stage motion. The gradual decrease in advancing CAs takes place over the saturation length $\lambda$ = 14 mm. This implies that the detachment of the grounded wire from the drop results in a decaying charge separation over the saturation length ($\lambda$) (Fig 4, Fig S7). Once the drop is saturated with charges and reaches its maximum potential, the charge separation behind the drop vanishes ($\sigma = 0$).

To detect the surface charge density gradient formed by the previous drop, we immediately slide a second grounded 5 µL drop along the same path. For the first 13.5 mm stage motion, we measure an advancing CA of 100° (Fig 4). The reduced advancing CA is due to an increase in surface energy by the deposited charges from the preceding grounded drop to insulated drop. Beyond 13.5 mm of stage motion, we observe a gradual increase in the advancing CA until 27 mm of stage motion. The stage position where the gradual increase in the advancing CA takes place is shifted by drop length, which in our case is 2.5 mm. Thereafter, the advancing CA remains constant at 113° for the rest of the stage motion. The gradual increase in the advancing CA of the grounded drop confirms the existence of surface charge density gradient. Regardless of the presence of a surface charge density gradient or drop charging state, we measure a constant receding CA of 83°. Therefore, spontaneous electrowetting and surface charge effects compensate each other ($\Delta\gamma_{SL} = \Delta\gamma_S$) at the receding CA during all stages of drop sliding.

## Conclusion

The drop charge develops an electric potential ~ 1 kV and leads to spontaneous electrowetting. Deposited charges behind the drop lead to surface charge effect and increase the surface-air interface energy ($\Delta\gamma_S$) ~ 10 mN/m. These two effects balance the receding CA irrespective of the drop charging state. Charge separation substantially influences the sliding contact angles, implying a significant contribution in the wetting dynamics of drops on surfaces.

# Acknowledgement

This work is supported by the German Research Foundation (DFG) within the framework Collaborative Research Centre 1194 ''Interaction of Transport and Wetting Processes'', Project-ID 265191195, subproject C07 (C. Hinduja, R. Berger, and H.-J. Butt) and A02b (A. Ratschow), the Priority Program 2171 ''Dynamic wetting of flexible, adaptive, and switchable surfaces'' (Grant No. BE 3286/6-1, B. Leibauer and R. Berger), the UGC-Junior research fellowship (award number-190510224296, S. Singh), the cooperation program of the Max Planck Society with the Indian Institutes for Science Education and Research (R. Chaurasia), and the European Research Council (ERC) under the European Union's Horizon 2020 research and innovation program (Grant No. 883631, N. Knorr, A. Ratschow, H.-J. Butt).

# Authors Contribution

C.H. conceived, designed, performed the experiments, and analyzed the data. B.L. prepared and characterized the samples, and performed the AFM experiments. C.H. and R.C. analyzed the contact angle data. N.K. performed Kelvin Probe experiments. A.D.R. suggested the principle of virtual work and noted the absence of viscous dissipation. S.S. & C.H. performed drop capacitance measurements. H.-J.B. proposed and derived the mathematical model. C.H., R.B., & H.-J.B. interpreted the data. C.H. and R.B. wrote the manuscript.

# Conflict of Interest

The authors declare no conflict of interest.